\documentclass{article}

\everymath{\rm}
\everydisplay{\rm}

\usepackage{graphicx}
\begin{document}
\begin{center}
{\bf \large  Measurement of $R_{LT}$ and $A_{TL}$ in the
$^4He(e,e'p)^3H $ Reaction at $p_{miss}$ of 130-300 MeV/c.}\\[0.2in]
K. A. Aniol,
M. B. Epstein$^1$, 
E. Gama$^a$,
D. J. Margaziotis\\
{\it Department of Physics and Astronomy, California State University, Los Angeles, Los Angeles, California 90032, USA}\\[0.1in]
W. Bertozzi,
J. P. Chen$^b$, 
D. Dale$^c$,
R. E. J. Florizone,
S. Gilad,
A. J. Sarty$^d$,
J. A. Templon$^e$,
S. P. Van Verst$^f$,
J. Zhao,
Z.-L. Zhou$^g$\\
{\it Laboratory for Nuclear Science, MIT, Cambridge, Massachusetts 02139,
USA}\\[0.1in]
P. Bartsch,
W. U. Boeglin$^h$,
R. Boehm,
M. O. Distler,
I. Ewald,
J. M. Friedrich$^i$,
J. M. Friedrich,
R. Geiges,
P. Jennewein,
M. Kahrau,
K. W. Krygier,
A. Liesenfeld,
H. Merkel,
K. Merle,
U. Muller,
R. Neuhausen,
E. A. J. M. Offermann$^j$,
Th. Pospischil,
G. Rosner$^k$,
H. Schmieden$^l$,
A. Wagner,\\
Th. Walcher\\
{\it Institut f\"{u}r Kernphysik, Universit\"{a}t Mainz, D-55099 Mainz, Germany}\\[0.1in]
M. Kuss$^m$,
A. Richter,
G.  Schrieder\\
{\it Institut f\"{u}r Kernphysik, Technische Universit\"{a}t Darmstadt, D-64289 Darmstadt,
Germany}\\[0.1in]
K. Bohinc,
M. Potokar,
S. \v{S}irca\\
{\it Institute "Jo\v{z}ef Stefan"and University of Ljubljana, SI-1001 Ljubljana, Slovenija}\\[0.1in]
J. M. Udias, Javier R. Vignote\\
{\it Departamento de Fisica Atomica, Molecular y Nuclear, Facultad de Ciencias Fisicas,
Universidad Complutense de Madrid, E-28040 Madrid, Spain}\\[0.1in]
R. Schiavilla\\
{\it Department of Physics, Old Dominion University, Norfolk, Virginia 23529, USA}\\[0.2in]

PACS numbers: 25.10+s, 25.30.RW, 21.45.+v, 27.10+h\\[0.2in]
\end{center}
{\bf Abstract}\\
We have measured the $^4He(e,e'p)^3H$ reaction at missing momenta of 130-300 MeV/c using the 
three-spectrometer facility at the Mainz microtron MAMI. Data were taken in perpendicular
kinematics to allow us to determine the response function $R_{LT}$ and the asymmetry term $A_{TL}$.
The data are compared to both relativistic and non-relativistic calculations.\\[0.2in]
{\bf 1. Introduction}\\
There have been many studies of the (e,e'p) reaction with the overall
objective of learning more about the detailed single nucleon distributions in nuclei. Of
special interest are investigations of few body systems and light nuclei 
since they are more amenable to more complete and detailed theoretical calculations. 
Interpretation of the experimental results in terms of single nucleon properties is most
readily done in the simple non-relativistic Plane Wave Impulse Approximation (PWIA).
However, it has been clear for sometime now that the simple non-relativistic PWIA
is not able to account for the observed cross sections. One has to understand
the nature of the competing processes, particularly in the more interesting kinematic regions of
high missing momenta. An important experimental tool in this regard is the response function
decomposition of the cross sections.   
In the one photon exchange approximation the cross section can be written as \cite{Pick} :\\[0.1in]
\begin{displaymath}
\frac { d^5 \sigma } { d \Omega_e d \Omega_p d E_e }
 = K ~\sigma_{Mott}[ \nu_L R _ L + \nu_T R_T + \nu_{ LT }
R_{ L T } cos \phi  + \nu_{ T T } R_{ T T } cos 2 \phi ]~~~(1)
\end{displaymath}\\[0.1in]
where $ \phi $ is the angle between the electron scattering
plane and the plane containing the momentum
transfer $ \vec q $ and the detected
proton, K is a kinematic factor $ \frac { \displaystyle m p _ p } { \displaystyle ( 2 \pi ) ^ 3 }$
(where $ p _ p$ is the momentum of the ejected proton),
$ \sigma  _ { Mott}$ is the Mott cross section, the $ \nu _ i $ are additional
kinematic factors and the $ R _ i$ and $R _ { i j }$ are the response
functions. The response functions describe the interaction
of the longitudinal, L, and transverse, T, polarization states of the
virtual photon with the nuclear charge and current.
Theoretical calculations suggest that each of
these response functions can exhibit selective sensitivity to particular reaction mechanisms
in the (e,e'p) process.
For example, past measurements of the $R _{LT}$ interference response function
have indicated that it is sensitive to relativistic effects and meson exchange currents (MEC).
Measurements of the $^2 H(e,e'p)n $ reaction \cite{Steen} have shown that it is
sensitive to inclusion of relativistic terms
in the nucleon current while a previous study \cite{Epst} of the
$^4He(e,e'p)^3H$ reaction indicated the need to include meson exchange terms in order to
reproduce both the cross sections and the $R_{LT} $ response function measured for missing momentum
$p_{miss}$~=~265 MeV/c.
Studies of the $ ^{16} O(e,e'p) ^{15} N $ reaction have suggested that the $R_{LT}$ response function is sensitive
to both MEC \cite {Ryck} and a fully relativistic treatment \cite {Gao} of the reaction. 

The $^4 He $ system is of particular interest since it is a tightly bound system
for which one  can obtain the nuclear wavefunction from microscopic calculations based on
realistic NN interactions. In this paper we report on a measurement of the $R_{LT} $ interference
response function in the $^4He(e,e'p)^3H $ reaction.  These particular measurements were motivated 
in part by the measurement of $ R_ { LT}$, performed at the MIT-Bates Laboratory,
for the $^4He(e,e'p)^3H $ reaction at missing momentum $p_{miss} = 265~MeV/c $, incident
electron energy, $E_o = 572.5~MeV$, momentum transfer, 
$ |\vec q|~= 360~MeV/c$, and energy transfer, $ \omega  = 200~MeV$ \cite{Epst}. 
The present experiment was  designed to provide measurements of $R _{LT} $ under similar kinematic conditions
but over a wider range of $p_{miss}$.\\[0.4in] 
{\bf 2. The Experiment}\\
The research reported here was done with the three-spectrometer facility at the Mainz 
microtron, MAMI, by the A1 collaboration. Data were taken at incident beam energies of 675 and 855
MeV and cover a range of $p_{miss}$ from 131 MeV/c to 300 MeV/c.  As was the case for the data of Ref.\cite{Epst}
$ \omega $ was set at approximately 200 MeV and kept relatively constant in an attempt to keep
final state interaction (FSI) effects constant for all
measurements and also to minimize FSI effects since proton-nucleus energies of around 200 MeV are near
the minimum of the $p-^4He$ optical model potential derived by Van Oers et al \cite{VanO}.

The experimental setup was almost identical to that described in Ref. \cite{Flori1} and Ref. \cite{Flori2}
except that for the present measurements the electron and proton spectrometers were
interchanged, i.e. Spectrometer A was used to detect protons and Spectrometer B was used to detect electrons. (Details
of the MAMI three spectrometer facility are described in Ref. \cite{Blom}.) Spectrometer C was used as
a luminosity monitor. The angular acceptance of Spectrometer B,
as defined by the data analysis cuts, was $ 2.29 ^o $ in the
horizontal direction and $ 7.45 ^ o $ in the vertical direction. The large angular acceptance of the
proton spectrometer, spectrometer A, allowed us to break each angular setting into three 
data points of angular acceptance of $ 4.30 ^ o $ (vertical acceptance is $11.46 ^o $).
The target consisted of cold $ ^ 4 He $ gas (T = 20-23 K and P = 5-10 atm) encapsulated
in an 8 cm diameter stainless steel quasi-spherical cell whose walls were $82~ \mu m$ thick \cite{Blom2}.

The target density was determined \cite{Flori2} by
measuring elastic scattering from $ ^ 4 He $ in Spectrometer B and normalizing the measured counts to
the elastic scattering data and form factor parametrization of Ref. \cite{Otter}.
Spectrometer C was configured to detect negatively charged particles and
data were taken in Spectrometer C during the elastic scattering
runs. For each beam energy the angle and momentum settings for Spectrometer C were kept fixed.
Thus, with appropriate cuts to eliminate background, the recorded counts in Spectrometer C, along with the measured
beam current, provided a measurement of the target density for any particular data run relative to what was
determined from the elastic scattering data runs.
The effective target length, as defined by the data analysis
cuts, varied depending on the particular kinematic settings between 5.2 to 6.0 cm.
The combined effective solid angles and target length of the two-spectrometer extended-target system was
determined using the Monte Carlo code AEEXB \cite{Off}.

Beam energies of 855 and 675 MeV were used with average
beam currents of $40~ \mu  A$.  Due to constraints imposed by the available beam energies and the geometry
of the experimental hall we were not able to keep $\vec q $ constant and $|\vec q|$ varied from 404.6 to 639.5 MeV/c.
These changes in $|\vec q| $ are responsible for the discontinuities that are
seen when the data are plotted as a function of $ p _ {miss} $. 

To determine the asymmetry term, $ A _ {TL} $, and the interference response function, $ R _{LT} $,
data were taken in perpendicular kinematics.
\begin{displaymath}
A_{ T L } ~=~ \frac{\sigma_f~-~\sigma_b}{\sigma_b~+~\sigma_f}~~~~~~~~~~~~~~~~~~~~~~~~~~~~~~~~~~~~~~~~~~~(2) 
\end{displaymath}
$ R _{LT} $ is defined in Equation 1.
For each value of $p _{miss} $ two measurements were made, the first with
the proton spectrometer set at an angle forward of  $ \vec q $ (closer to the beam direction) to
meaure $ \sigma_f $,
and the second with the proton spectrometer set at an equal angle backward of $ \vec q $ ( further
away from the beam direction) to measure $ \sigma_b$. Under these conditions the central value of $ p _{miss}$ is the
same for both measurements. 

Data were taken at one set of angles at the 855 MeV beam energy and
three different sets of angles at the 675 MeV
beam energy. After breaking up the horizontal angular acceptance of the proton spectrometer into
three angular regions, each angular setting yielded three data points,
with each data point spanning a total width in $p_{miss}$ between 25 to 58 MeV/c.
The kinematics for these data are shown in Table 1. To determine $ A _ { TL } $ and $ R _ { LT } $ 
the same cuts were put on $|\vec q| $, $ \omega $, and $p_{miss}$ for each data point of the pair
of angular settings used. 

The combined energy resolution of spectrometers A and B along with the approximately 100 keV spread in the beam energy
resulted in an $E _ {miss}$ spectrum in which the two-body breakup peak had a FWHM of approximately 700 keV. (See Figure 1.)
\begin{figure}[!hbtp]
\includegraphics[width=3.5in]{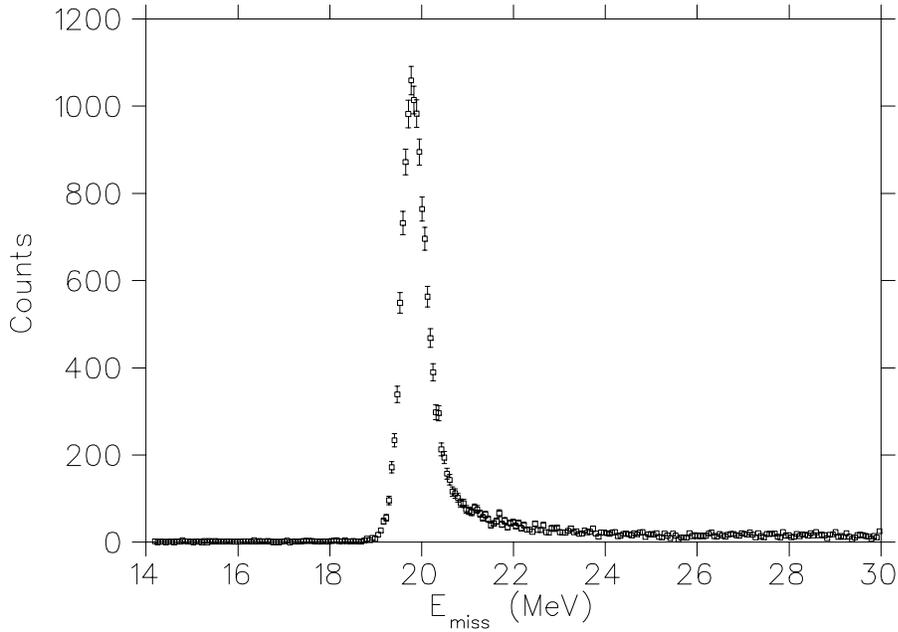}
\caption{$^4He(e,e'p)^3H.~~E_ { miss }$ spectrum for $p_ {miss}= 150~MeV/c$. The three body breakup threshold is at $E_{miss}=26.1~MeV $. }
\end{figure}
This was more than sufficient to provide clean separation between the two-body breakup peak and the threshold
for the three-body continuum.  Total combined systematic errors for individual cross section measurements
were approximately 6\%.  Data are presented here only for the $^4He(e,e'p)^3H$ reaction, that is only
for the two-body final state $E_ {miss}$ peak.  The data were corrected for radiative losses due to
internal and external bremsstrahlung processes \cite{brem1}, \cite{brem2}.\\[0.2in]
{\bf 3. Results}\\
The cross section data for proton angles forward of $\vec q $ ($\sigma_f$) are shown in Figure 2 and those for angles back of
$\vec q $ ($\sigma_b $) are shown in Figure 3. Only statistical errors, which range between 1\% to 7\% , are plotted and these 
are almost always smaller than the size of the data points. The cross section data are tabulated in Table 1.
\begin{figure}[hbt]
\includegraphics[width=4.0in]{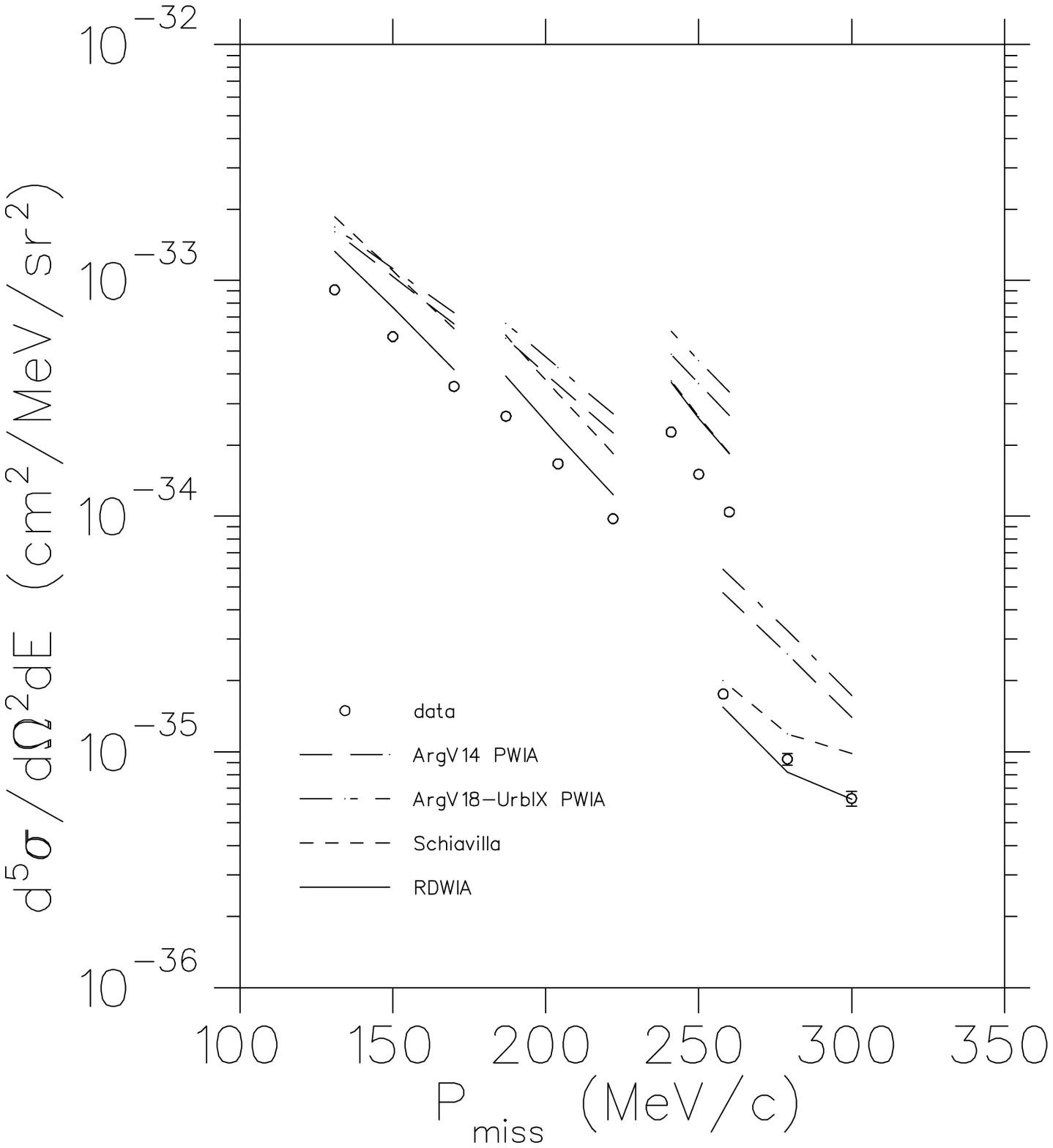}
\caption{$^4He(e,e'p)^3H.~~$Cross sections for proton angles forward of q}
\end{figure}
\begin{figure}[hbt]
\includegraphics[width=4.0in]{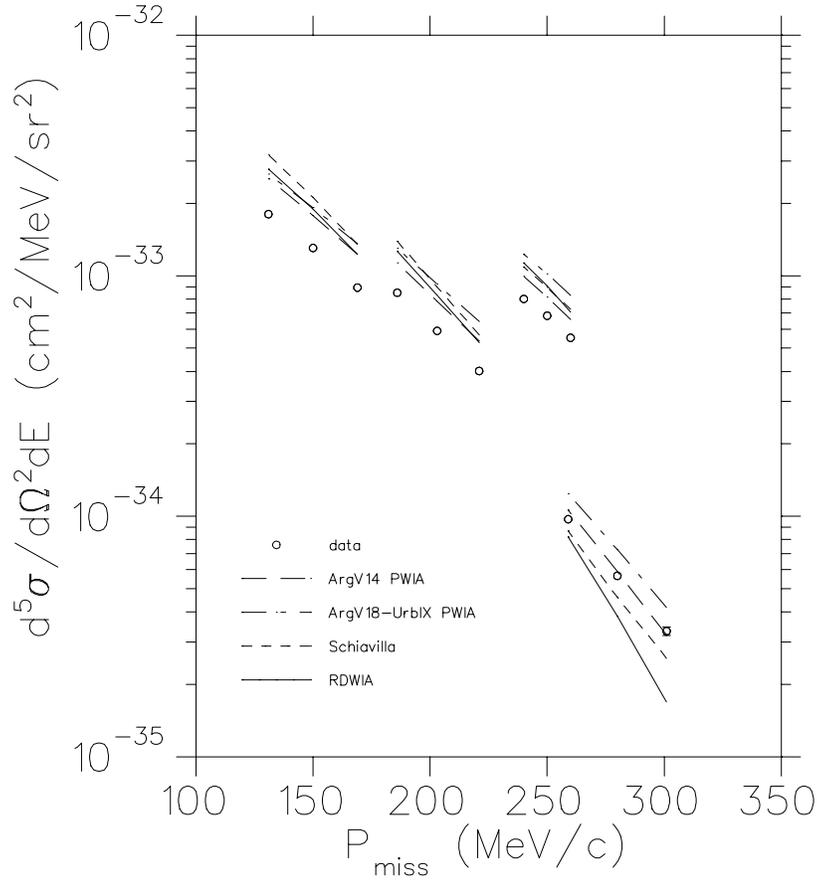}
\caption{$^4He(e,e'p)^3H.~~$Cross sections for proton angles backward of q  }
\end{figure}
Also shown in Figures 2 and 3 are four different theoretical predictions. We show two 
PWIA calculations, one using a $^4He $ wavefunction derived from the Argonne V14 nucleon-nucleon
potential \cite{Arg14} and the other using a $^4 He $ wavefunction derived from the Argonne V18
nucleon-nucleon potential and the Urbana IX three nucleon potential \cite{ArgV18}.
The calculations labeled Schiavilla incorporate the orthonormal-correlated states method described by Schiavilla 
\cite{Schiav} and include the effects of short range correlations,
orthogonality corrections, final state interactions, and two-body charge and current operators.

The curves labeled RDWIA are fully relativistic distorted wave
impulse approximation (RDWIA) calculations by Udias using the Madrid code
\cite{Udias}. This calculation uses the same ingredients as in the
corresponding ones in ref.\cite{Diet} that were compared to the transferred
polarization ratio data at similar energies of the ejected proton as in
this experiment.  Namely: a) a relativistic mean field wave function fit
to reproduce the rms radius and binding energy of $^4He$ which
reproduces the momentum distributions from the $^4He$ data in Ref.\cite{Flori2}. b)
an optical potential obtained by folding a density-dependent empirical
effective p-N interaction (EEI) \cite{Kelly} with the measured charge density for
tritium. Here we use the same potential as in Ref. \cite{Diet}, derived from
parameters that were adjusted by Kelly to fit proton scattering data from
$^4He$, obtaining a better fit to the proton elastic scattering data in this
nucleus than any previous optical potential. c) The CC1 current operator.

The PWIA calculations were averaged over the finite acceptances of the spectrometers while the full Schiavilla 
and the RDWIA calculations of Udias were done only at the central point of these acceptances. Using the Argonne V14 PWIA
calculation we compared the effects of acceptance averaging these calculations. The difference between the point 
and acceptance averaged calculations were typically less than 8 \%.

Figure 4 shows the asymmetry term $ A _ { T L } $ 
and Figure 5 shows $ R _ { L T} $.
\begin{figure}[!hbtp]
\includegraphics[width=4.0in]{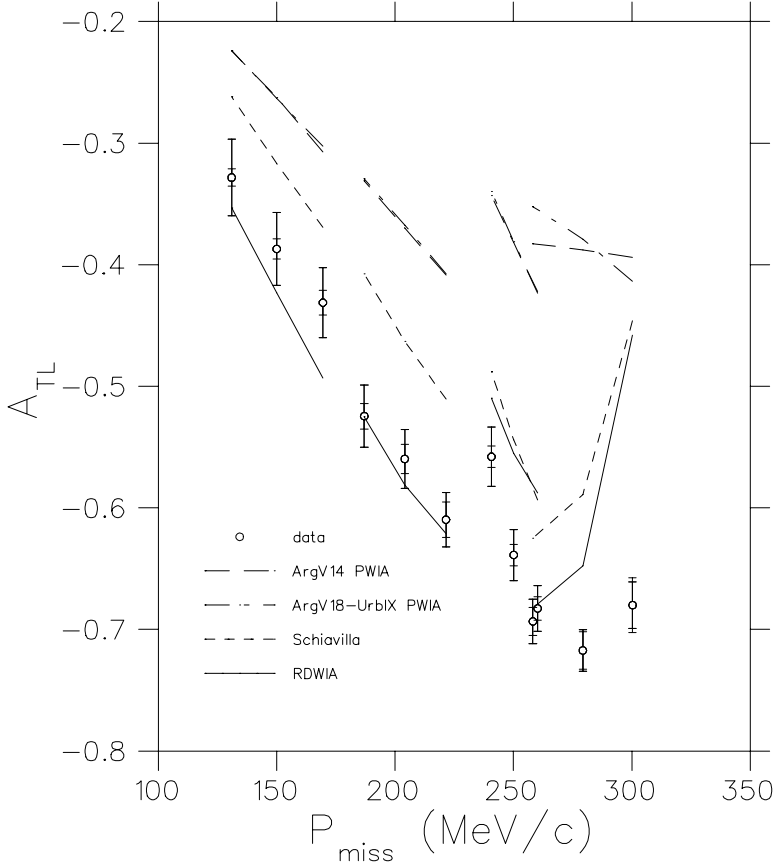}
\caption{$^4He(e,e'p)^3H.~~$The transverse-longitudinal asymmetry $A_ { TL } $ }
\end{figure}
\begin{figure}[!hbtp]
\includegraphics[width=4.0in]{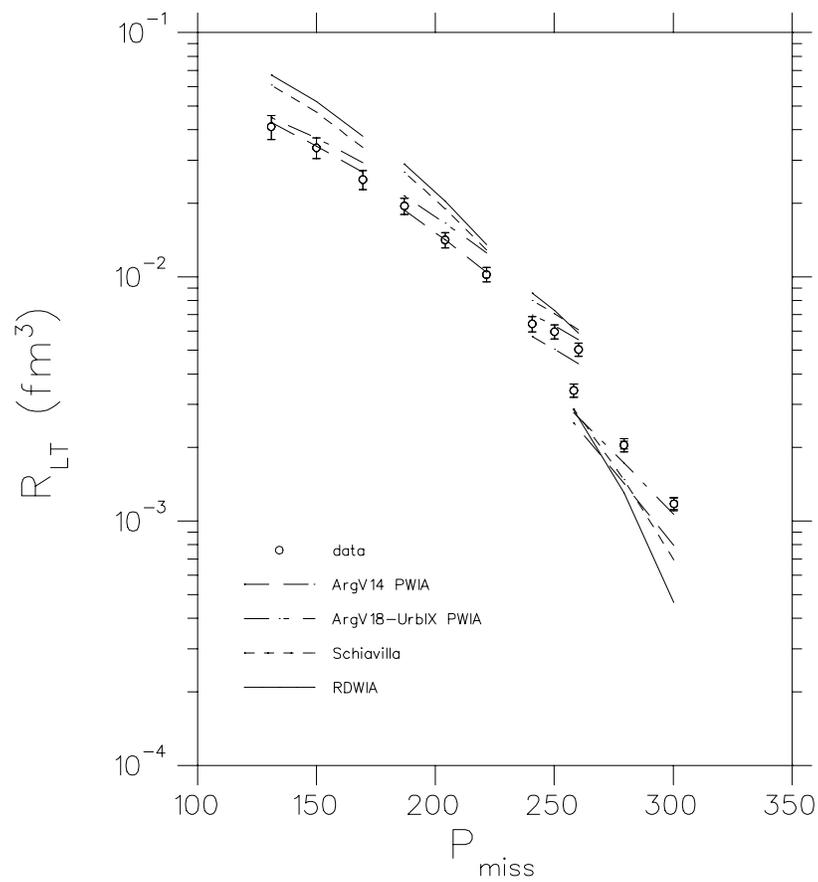}
\caption{$^4He(e,e'p)^3H.~~$The longitudinal-transverse response function $R_{ LT } $ }
\end{figure}
In these figures the two sets of error bars represent the statistical and systematic
errors separately. Except for $ A _ { T L } $ at $ p_{miss}  =  300~MeV/c$ the systematic errors are always larger
than the random errors.  
The theoretical predictions shown in Figures 4 and 5 are derived from the cross section 
calculations shown in Figures 2 and 3. To investigate the effects of MEC on $ R_ { LT} $ and $ A _ { TL } $
the Schiavilla calculations were also run without the MEC term but with everything else included. Figures 6 and 7
show these results.
\begin{figure}[!hbtp]
\includegraphics[width=4.0in]{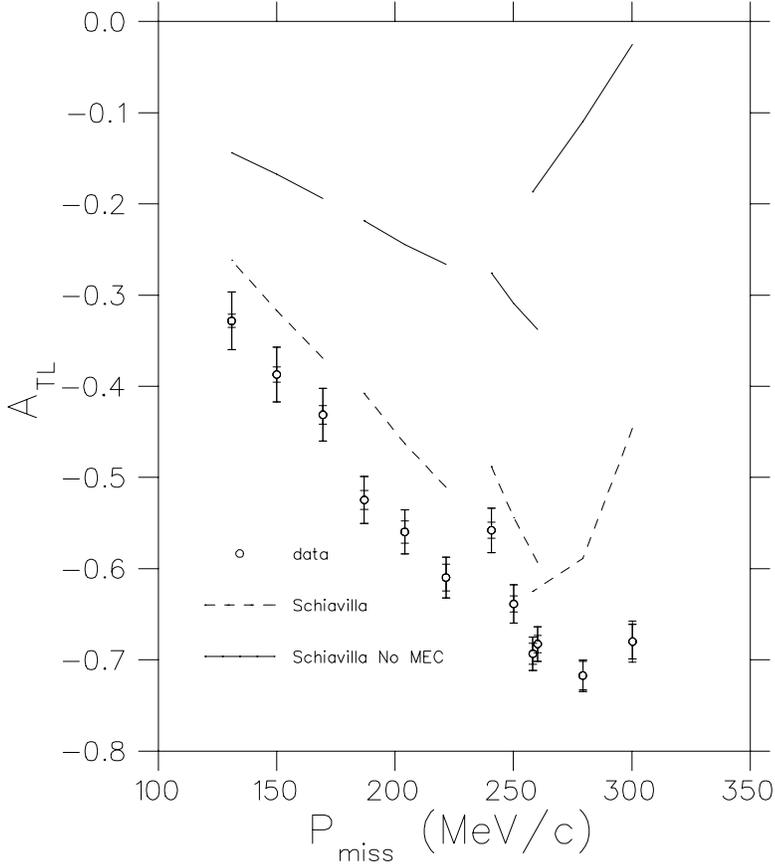}
\caption{$^4He(e,e'p)^3H.~~A_ {TL}$ Schiavilla calculations with and without MEC terms }
\end{figure}
\begin{figure}[!hbtp]
\includegraphics[width=4.0in]{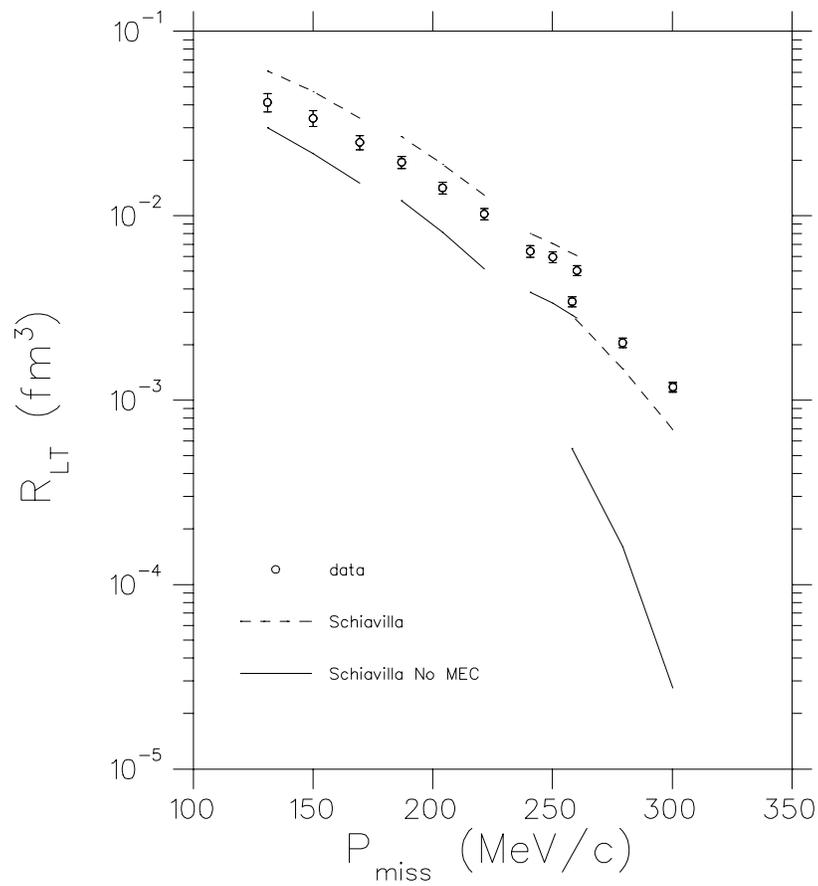}
\caption{$^4He(e,e'p)^3H.~~R_{ LT } $  Schiavilla calculations with and without MEC terms }
\end{figure}

In comparing theory with the cross section, $ R _ { LT } $, and 
$ A_ { TL } $  data it is difficult to draw any firm conclusions. There is some indication from  $ A _ { TL } $ 
that the RDWIA calculations of Udias or the full calculation of Schiavilla is necessary to produce the
qualitative shape of the data.  The inclusion of MEC terms in the Schiavilla calculation also substantially
improves agreement with the $ R _ { LT } $ data as was seen in the earlier MIT-Bates results
\cite{Epst}.  However it is less clear which predictions produce the best fits to the
cross section data. The full calculation of Schiavilla or the
RDWIA calculations of Udias agree best with the forward angle cross section data while the
simple PWIA calculation seems to do better for the back angle cross sections.

We note that the data measured at 855 MeV have Bjorken x = 1.08 and the largest value of $Q^{2}$.
From the point of view of the electron kinematics the 855 MeV data should have
been the best case for quasi-elastic scattering ( x near 1 ) with the smallest
contribution from meson exchange currents. In contrast, at 675 MeV and $\theta=50^o$, x= 0.58
which should imply a greater role for meson exchange currents.
However Figures 6 and 7 indicate that, at least in the context of the Schiavilla calculation, meson exchange
currents are more important for the 855 MeV data than for the 675 MeV data.
Overall these data suggest that the inclusion of MEC terms and/or a fully relativistic calculation is
necessary but at this time we are unable to conclude which effect is dominant.\\[0.2in]
{\bf 4. Summary}\\
We have taken data for the $^4He(e,e'p)^3H $ reaction in perpendicular kinematics that allowed us 
to determine $ R _ { LT} $ and $ A _ {TL} $ for $p _ {miss} $ from 130 to 300 MeV/c. Reasonable 
agreement with theory appears to require inclusion of final state interaction effects and
the addition of meson exchange reaction terms and/or a fully relativistic treatment of the entire
reaction. Additional data might clarify this situation. In this regard there is a proposal under
review for a new series of measurements of the $^4He(e,e'p)^3H $ reaction to be done
at Jefferson Lab.\\[0.2in]

We wish to thank the MAMI staff for their support during these experiments. This work was supported by
the Deutsche Forschungsgemeinshaft (SFB 201) and Ri 242/15-2, 
by the U.S. National Science Foundation, and by the U.S. Department of Energy.

\clearpage

\begin{table}[h]
\begin{center}
\begin{tabular}{|c|c|c|c|c|c|c|c|c|}
\hline

$E_o $ & $\mathrm{\theta _e}$ & q & $\mathrm{\theta _p}$ & $\mathrm{ \omega~}$ & $ p_p$ & $p_{miss}$& $d^5\sigma$/$d\Omega^2dE$ &\% error \\
MeV & deg.  & MeV/c & deg.  & MeV & MeV/c & MeV/c & $cm^2$/MeV/$sr^2$ &stat.   \\
\hline
855.11 & 47.70 & 639 & 22.86 & 185.11 & 549.95 & 300.3 & $6.34~x~10^{-36}$ & 7 \\
855.11 & 47.70 & 639 & 25.01 & 185.11 & 554.19 & 279.3 & $9.31~x~10^{-36}$ & 6 \\
855.11 & 47.70 & 639 & 27.16 & 185.11 & 558.14 & 258.2 & $1.76~x~10^{-35}$ & 4 \\
855.11 & 47.70 & 639 & 78.85 & 185.11 & 549.76 & 301.2 & $3.33~x~10^{-35}$ & 4 \\
855.11 & 47.70 & 639 & 76.70 & 185.11 & 554.01 & 280.3 & $5.65~x~10^{-35}$ & 3 \\
855.11 & 47.70 & 639 & 74.55 & 185.11 & 557.97 & 259.1 & $9.72~x~10^{-35}$ & 2 \\
\hline
675.11 & 55.00 & 559 & 27.50 & 208.11 & 614.26 & 169.5 & $3.55~x~10^{-34}$ & 2 \\
675.11 & 55.00 & 559 & 29.65 & 208.11 & 616.28 & 150.0 & $5.77~x~10^{-34}$ & 2 \\
675.11 & 55.00 & 559 & 31.80 & 208.11 & 618.02 & 131.0 & $9.12~x~10^{-34}$ & 1 \\
675.11 & 55.00 & 559 & 58.93 & 208.11 & 614.28 & 169.4 & $8.93~x~10^{-34}$ & 1 \\
675.11 & 55.00 & 559 & 56.78 & 208.11 & 616.30 & 149.9 & $1.31~x~10^{-33}$ & 1 \\
675.11 & 55.00 & 559 & 54.63 & 208.11 & 618.03 & 130.8 & $1.80~x~10^{-33}$ & 1 \\
\hline
675.11 & 36.00 & 404 & 22.86 & 208.11 & 601.55 & 260.2 & $1.04~x~10^{-34}$ & 3 \\
675.11 & 36.00 & 404 & 25.01 & 208.11 & 603.24 & 250.1 & $1.50~x~10^{-34}$ & 3 \\
675.11 & 36.00 & 404 & 27.16 & 208.11 & 604.75 & 240.7 & $2.27~x~10^{-34}$ & 2 \\
675.11 & 36.00 & 404 & 62.56 & 208.11 & 601.60 & 259.9 & $5.52~x~10^{-34}$ & 2 \\
675.11 & 36.00 & 404 & 60.41 & 208.11 & 603.28 & 249.8 & $6.82~x~10^{-34}$ & 2 \\
675.11 & 36.00 & 404 & 58.26 & 208.11 & 604.79 & 240.4 & $8.01~x~10^{-34}$ & 2 \\
\hline
675.11 & 50.00 & 518 & 22.86 & 208.11 & 607.65 & 221.5 & $9.74~x~10^{-35}$ & 4 \\
675.11 & 50.00 & 518 & 25.01 & 208.11 & 610.07 & 204.1 & $1.67~x~10^{-34}$ & 3 \\
675.11 & 50.00 & 518 & 27.16 & 208.11 & 612.24 & 187.0 & $2.65~x~10^{-34}$ & 3 \\
675.11 & 50.00 & 518 & 64.41 & 208.11 & 607.74 & 220.9 & $4.02~x~10^{-34}$ & 2 \\
675.11 & 50.00 & 518 & 62.26 & 208.11 & 610.15 & 203.4 & $5.90~x~10^{-34}$ & 2 \\
675.11 & 50.00 & 518 & 60.11 & 208.11 & 612.32 & 186.4 & $8.50~x~10^{-34}$ & 2 \\
\hline
\end{tabular}
\end{center}
\caption{ Kinematics (Central Angles and Momenta) and Cross Sections}
\end{table}
\bigskip 
\bigskip
\begin{flushleft}
$^1$ e-mail address: epstein@calstatela.edu\\[0.1in]
$^{(a)}$ Present address: Jet Propulsion Laboratory, Pasadena CA, 91109, USA\\
$^{(b)}$ Present address: TJNAF, Newport News, VA 23606, USA\\
$^{(c)}$ Present address: University of Kentucky, Lexington, KY, 40506, USA\\
$^{(d)}$ Present address: Saint Mary's University, NS B3H3C3, Canada\\
$^{(e)}$ Present address: NIKHEF, Amsterdam, The Netherlands\\
$^{(f)}$ Present address: Washington State Department of Health, Olympia, WA 98504, USA\\
$^{(g)}$ Present address: Schlumberger-Doll Research, Ridgefield, CT 06877, USA\\
$^{(h)}$ Present address: Florida International University, Miami, FL, 33199, USA\\
$^{(i)}$ Present address: Physik Department E18, Technische Universit\"{a}t M\"{u}nchen, D-85748 Garching bei M\"{u}nchen, Germany\\
$^{(j)}$ Present address: Renaissance Technologies Corp., Setauket, New York 11733, USA\\
$^{(k)}$ Present address: University of Glasgow, G12 8QQ, Scotland, UK\\
$^{(l)}$ Present address: Universit\"{a}t Bonn, 53115 Bonn, Germany\\
$^{(m)}$ Present address: INFN, Sezione di Pisa, Pisa Italy\\
\end{flushleft}

\end{document}